\documentclass[aps,prb,superscriptaddress,twocolumn]{revtex4-1}

\usepackage{lineno,hyperref}
\usepackage{graphicx}
\usepackage{bmpsize}
\usepackage{amsfonts}

\newcommand{\ben}{\begin{eqnarray}}
\newcommand{\een}{\end{eqnarray}}

\newcommand{\bef}{\begin{figure}[h!bt]\centering}
\newcommand{\eef}{\end{figure}}
\newcommand{\bet}{\begin{table}[hbt]\centering}
\newcommand{\eet}{\end{table}}

\begin{document}
\title{Magneto-transport properties of the ``hydrogen atom'' nodal-line semimetal candidates CaTX (T=Ag, Cd; X=As, Ge)}
\author{Eve Emmanouilidou}
\email{These authors contribute equally.}
\affiliation{Department of Physics and Astronomy and California NanoSystems Institute, University of California, Los Angeles, CA 90095, USA}
\author{Bing Shen}
\email{These authors contribute equally.}
\affiliation{Department of Physics and Astronomy and California NanoSystems Institute, University of California, Los Angeles, CA 90095, USA}
\author{Xiaoyu Deng}
\affiliation{Department of Physics and Astronomy, Rutgers University, Piscataway, NJ 08854, USA}
\author{Tay-Rong Chang}
\affiliation {Department of Physics, National Cheng Kung University, Tainan, 701, Taiwan}
\author{Aoshuang Shi}
\affiliation{Department of Physics and Astronomy and California NanoSystems Institute, University of California, Los Angeles, CA 90095, USA}

\author{Gabriel Kotliar}
\affiliation{Department of Physics and Astronomy, Rutgers University, Piscataway, NJ 08854, USA}
\author{Su-Yang Xu}
\affiliation{Department of Physics, MIT, Cambridge, MA 02139, USA}
\author{Ni Ni}
\email{Corresponding author: nini@physics.ucla.edu}
\affiliation {Department of Physics and Astronomy and California NanoSystems Institute, University of California, Los Angeles}
\begin{abstract}
Topological semimetals are characterized by protected crossings between conduction and valence bands. These materials have recently attracted significant interest because of the deep connections to high-energy physics, the novel topological surface states, and the unusual transport phenomena. While Dirac and Weyl semimetals have been extensively studied, the nodal-line semimetal remains largely unexplored due to the lack of an ideal material platform. In this paper, we report the magneto-transport properties of two nodal-line semimetal candidates CaAgAs and CaCdGe. First, our single crystalline CaAgAs supports the first ``hydrogen atom'' nodal-line semimetal, where only the topological nodal-line is present at the Fermi level. Second, our CaCdGe sample provides an ideal platform to perform comparative studies because it features the same topological nodal line but has a more complicated Fermiology with irrelevant Fermi pockets. As a result, the magneto-resistance of our CaCdGe sample is more than 100 times larger than that of CaAgAs. Through our systematic magneto-transport and first-principles band structure calculations, we show that our CaTX compounds can be used to study, isolate, and control the novel topological nodal-line physics in real materials.\end{abstract}
\pacs{}
\date{\today}
\maketitle

\section{Introduction}
Understanding the nontrivial topological properties in electronic band structures has become one of the central themes in condensed matter physics and materials science. Following the discovery of the 2D quantum Hall effect, the quantum spin Hall effect and 3D topological insulators \cite{david, bi2se3, PbTe, TlBiSe2}, interest has recently shifted toward realizing topological physics in gapless systems, i.e., topological semimetals. Topological semimetals are characterized by robust bulk band crossing points and the associated topological boundary states. They can be characterized by the dimension of the band crossing in momentum space. Two prominent examples of topological semimetals with 0D band crossing points are the Dirac and Weyl semimetals. Their realizations in Na$_3$Bi \cite{nb3bi-yulin, chiralna3bi}, Cd$_3$As$_2$ \cite{ cd3as2-cava, cd3as2-zahid}, TaAs \cite{taas-zahid, taas-hongding, nbp-felsa} and related materials have attracted enormous interest worldwide.

The other type of topological semimetal is the nodal-line semimetal whose conduction and valence bands cross to form a 1D closed loop in momentum space. The nodal-line semimetal differs from the Weyl semimetal in three aspects : (1) the bulk Fermi surface is 1D in nodal-line  and 0D in Dirac/Weyl semimetals ; (2) the density of states near the nodal touchings is proportional to $|E-E_{\textrm{F}}|^2$ in nodal-line and $|E-E_{\textrm{F}}|$ in Weyl semimetals ; (3) on the surface, the nodal-lines are ``stitched together" by a ``drumhead'' surface state, while Weyl nodes are connected by Fermi arc surface states. These unique properties of nodal-line semimetals make new physics accessible. For example, the weak dispersion of the drumhead surface states leads to a large density of states near the Fermi level. Therefore, possible interaction-induced instabilities on the surface of nodal-line semimetals have been widely discussed in theory \cite{superconducting-weyl-loops, instabilities, rahul}.
\begin{figure*}[t]
\includegraphics[width=160mm]{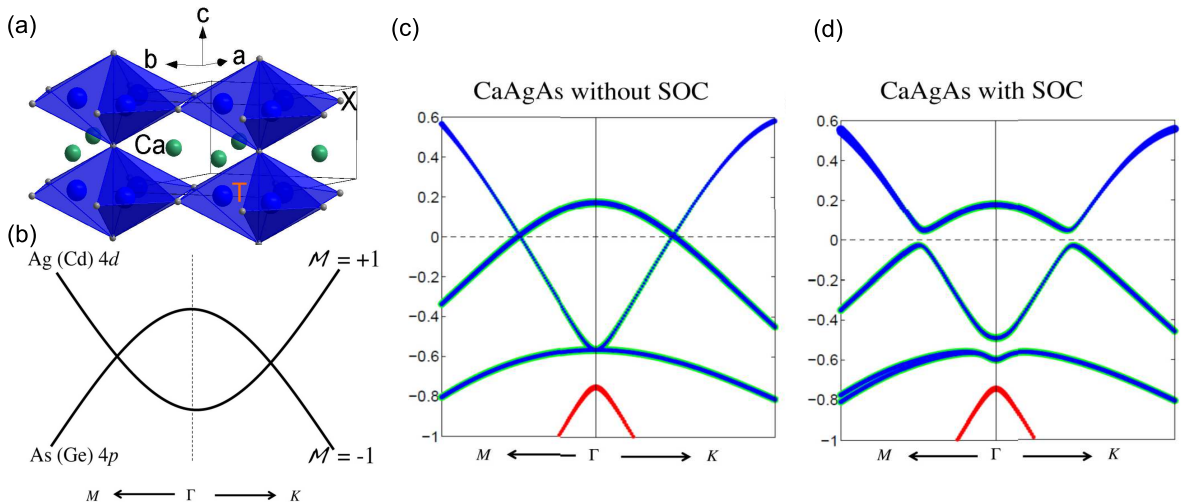}
\caption{(a) Crystal structure of CaAgAs and CaCdGe. TX$_4$ octahedra are shown in blue. Ca Atoms are shown in green. (b)-(c): Mirror symmetry protected nodal-lines in CaAgAs and CaCdGe: ({b}) Schematic of a band structure diagram for the nodal-line feature in CaAgAs and CaCdGe. The conduction and valence bands are made up of the Ag(Cd) $4d$ and As(Ge) $4p$ orbitals respectively. The band crossings near the $\Gamma$ point are protected because these two bands have opposite mirror eigenvalues. ({c,d}) First-principles calculated band structures of CaAgAs near the $\Gamma$ point with (d) and without (c) SOC.}
\label{Fig0}
\end{figure*}

Despite intense theoretical interest \cite{graphene, ca3p2, LaN, cu3xy, bavs3,pbtase2,zrsix,sriro3}, there haven't been many experimental studies on nodal-line semimetals. A major reason is the lack of ideal nodal-line semimetal materials, where only the nodal lines are present at the Fermi level with minimal irrelevant Fermi pockets. Such a ``hydrogen atom'' nodal-line semimetal is crucial for isolating the spectroscopic and transport signals from the nontrivial nodal lines from trivial states. For example, even before the discovery of TaAs, Fe, an elemental ferromagnetic metal, was known to have hundreds of Weyl nodes in its band structure \cite{Fe} . However, Fe is not an ideal platform to study Weyl physics because its complicated Fermi surface is dominated by irrelevant (non-Weyl) trivial pockets. In fact, because Weyl nodes are symmetry allowed when time-reversal or inversion symmetry is broken, they are likely to exist in the band structure of most ferromagnetic or non-centrosymmetric compounds. In this sense, the key is to identify a material where the topological band crossings (Dirac nodes, Weyl nodes, or nodal lines) are the dominant features at the Fermi level. Since such a platform is accessible in Dirac and Weyl semimetals (Cd$_3$As$_2$, Na$_3$Bi and TaAs), these topological materials have been under extensive study and novel physics has been discovered. By contrast, in nodal line semimetals, experimental work has been focused on PbTaSe$_2$ and ZrSiX (X=S, Se, Te) \cite{leslie, hu}. However, it is well-understood that both compounds have a quite complex band structure where multiple irrelevant pockets coexist with the nodal lines at the Fermi level.

The noncentrosymmetric CaAgX (X=As, P) compound crystalizing in the P-62m space group \cite{caagasstruc} was recently proposed to be a ``hydrogen atom'' nodal-line semimetal where two non-trivial bulk bands touch along a line and no trivial bands exist at the FL \cite{caagasdirac}. The crystal structure is shown in Fig. 1(a). It consists of a 3D-network of edge and corner sharing AgAs$_4$ tetrahedra. A recent study on polycrystalline CaAgX (X=As, P) revealed that it is a low carrier density metal and that CaAgAs is a more promising candidate than CaAgP for the purpose of studying the nodal lines \cite{caagaspowder}. Compared to polycrystals, single crystals are superior for transport studies and surface sensitive measurements. Until now however, no studies of single crystalline CaAgAs have been reported. In this paper, we report the magneto-transport properties of CaAgAs and its sister compound CaCdGe. CaCdGe also crystalizes in the P-62m space group \cite{CaCdGe}, featuring the same topological nodal line but with a much more complicated Fermiology, providing an ideal platform to perform comparative studies. We show that such a comparative study sheds light on a novel transport phenomenon prominently observed in many topological semimetals, the giant magnetoresistance.

\section{Experimental Methods}

CaAgAs crystals were grown with AgAs flux while CaCdGe crystals were grown using Cd flux. Ca granules and AgAs powder were mixed at a molar ratio of Ca:AgAs=1:4; Ca, Ge and Cd granules were mixed at a ratio of Ca:Ge:Cd=1:1:47. The materials were loaded into an alumina crucible and then sealed inside a quartz tube under 1/3 ATM of Ar. For CaAgAs, the ampule was heated to $1100^\circ$C, kept at that temperature for 3 hours, and cooled to $750^\circ$C at a rate of $3^\circ$C/hour. For CaCdGe, the ampule was heated to $800^\circ$C or $1000^\circ$C, dwelled for 3 hours and cooled to $400^\circ$C at a rate of $3^\circ$C/min. For CaCdGe, to compensate for the vaporized Cd in the growths which went up to $1000^\circ$C, 25\% extra Cd was added. These growths yielded much larger single crystals while the growths going up to $800^\circ$C yielded smaller needle-like single crystals with higher residual resistivity ratio (RRR). In both cases, the single crystals were separated from the flux by centrifuging.

X-ray diffraction measurements were performed using a PANalytical Empyrean (Cu K$\alpha$ radiation) diffractometer. Magnetotransport measurements were performed using a Quantum Design Physical Property Measurement System (QD PPMS Dynacool). The excitation current $I$ was set to 1-2 mA and along the $c$ axis, and magnetic fields up to 9 T were applied. We used the standard four- and six- probe techniques to measure the electrical resistivity $\rho_{xx}$ and Hall resistivity $\rho_{yx}$ with $\rho_{xx}(B)=\frac{\rho_{xx}(B)+\rho_{xx}(-B)}{2}$ and $\rho_{yx} = \frac{\rho_{yx}(B)-\rho_{yx}(-B)}{2}$ respectively. The electronic structure of CaCdGe was studied using first principles calculations based on density functional theory and the full-potential linear augmented plane wave method as implemented in the Wien2k package \cite{wien2k, PBE}. Spin orbit coupling (SOC) was taken into account in the calculation. The local density approximation (LDA) and the Tran-Blaha modified Becke-Johnson (MBJ) \cite {MBJ} exchange potential were used in our calculations.

\begin{figure}
  \includegraphics[width=3.4in]{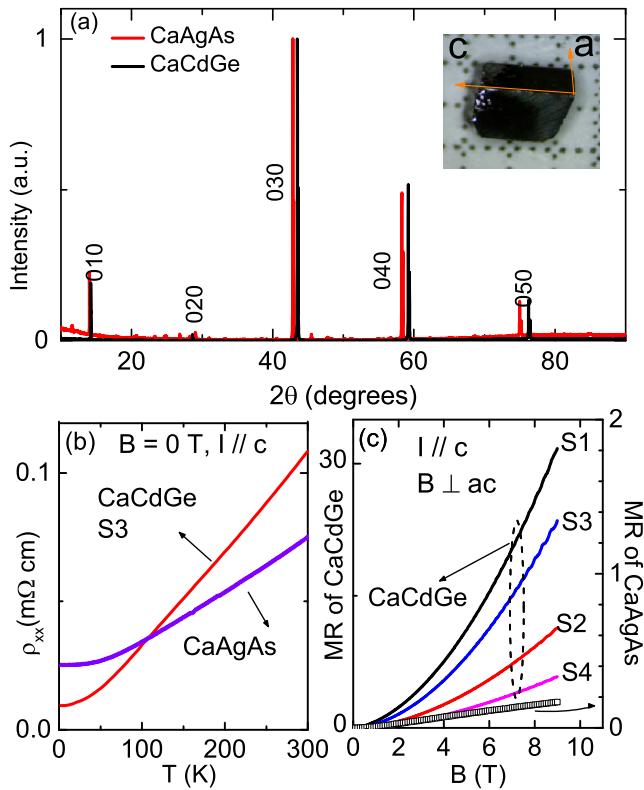}
  \caption{(a) The x-ray diffraction patterns of the as-grown $ac$ plane of CaCdGe and CaAgAs. Inset: CaCdGe single crystal on a mm scale. The crystalline directions are shown. (b) Temperature dependence of the electrical resistivity $\rho_{xx}$ for CaCdGe and CaAgAs at $B=0$ T with $I//c$. (c) MR of CaCdGe and CaAgAs single crystals at $T $= 2 K with $I//c$ and $B \perp ac$. }
  \label{fig1}
\end{figure}

\section{Experimental Results and Discussion}
The valence analysis (Ca$^{2+}$, Ag$^{1+}$, As$^{3-}$ for CaAgAs and Ca$^{2+}$, Cd$^{2+}$, Ge$^{4-}$ for CaCdGe) suggests a semimetal/semiconductor ground state for both compounds, which is confirmed by our band structure calculations. As shown in Fig.~\ref{Fig0}(c), the conduction and valence bands cross in the absence of SOC. The band crossing forms a 1D loop (a nodal-line) that encloses the $\Gamma$ point.

We now move on to explain the existence of this nodal line near the Fermi level. Our calculation shows that the lowest conduction and valence bands are made up of the Ag(Cd) $4d$ and As(Ge) $4p$ orbitals respectively. At the $\Gamma$ point, a band inversion takes place as the top of the valence (As $4p$) band moves above the bottom of the conduction (Ag $4d$) band. Furthermore, because the $z=0$ plane of the crystal is a mirror plane (Fig. 1(b)), the electron states on the $k_z=0$ plane must be eigenstates of the mirror operation $\mathcal{M}_z$, which takes $z$ to $-z$. Interestingly, the Ag $4d$ conduction band and the As $4p$ valence band have opposite mirror eigenvalues (Fig.~\ref{Fig0}(b)). This fact prevents them from hybridizing, leading to a nodal-line on the $k_z=0$ plane enclosing the $\Gamma$ point  (Fig.~\ref{Fig0}(b)). Upon the inclusion of SOC, a gap of less than 20 meV is opened  (Fig.~\ref{Fig0}(d)). Therefore, both CaAgAs and CaCdGe compounds are nodal-line semimetals when SOC is neglected, which is the same as many other nodal-line candidates including graphene networks, Ca$_3$P$_2$, LaN, Cu$_3$(Pd,Zn)N, and ZrSiX.
\begin{figure}
 \includegraphics[width=3.4in]{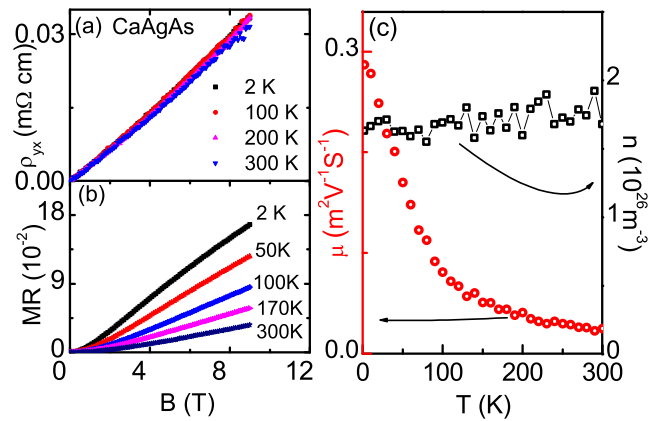}
  \caption{CaAgAs single crystal with $ I // c$ and B $\perp ac$: (a) Hall resistivity $\rho_{yx}$. (b) Field dependent transverse MR. (c) Temperature dependent carrier density and mobility. }
 \label{fig:}
\end{figure}

Based on the Rietveld refinement of the powder x-ray diffraction data, the lattice parameters are $a=b=7.3056(1)$\textrm{\AA}, $c=$ 4.4785(1)\textrm{\AA} for CaCdGe, and $a=b=7.2041(1)$\textrm{\AA}, $c=4.2699(1)$\textrm{\AA} for CaAgAs. As a representative, Fig. 2(a) shows the x-ray diffraction patterns of the as-grown $ac$ plane of CaCdGe and CaAgAs while the inset of Fig. 2(a) shows the sample orientation of a CaCdGe single crystal with the as-grown rectangular surface as the $ac$ plane and the hexagonal as-grown cross section as the $ab$ plane. CaCdGe and CaAgAs both demonstrate metallic behavior, as can be seen in Fig. 2(b). The RRR is 3 with a residual resistivity $\rho_0$ of $25~\rm{\mu \Omega~cm}$ for CaAgAs while it is 12 with a $\rho_0$ of $9~\rm {\mu \Omega~cm}$ for CaCdGe. Figure 2(c) shows the transverse magnetoresistance (MR) of four CaCdGe samples, and one representative CaAgAs sample. CaCdGe exhibits large, quadratic-like MR without a sign of saturation up to 9 T. S1 in particular, has an MR around 3200$\%$ at 2 K under 9 T. This behavior is reminiscent of the extremely large MR recently observed in materials such as Weyl semimetal TaAs, the type II-Weyl semimetal WTe$_2$, Dirac semimetal Cd$_3$As$_2$ and the weak topological insulator NbAs$_2$ \cite{taasjia, taaschen, WTe2, transcd3as2, nbas2}. In sharp contrast to the giant quadratic MR of CaCdGe, the MR of CaAgAs only goes up to about 18$\%$ and most notably has a non-quadratic character, which agrees with the recent work on polycrystalline CaAgAs \cite{caagaspowder}.

Figure 3(a) shows the Hall resistivity $\rho_{yx}$ for CaAgAs, which appears to be positive. It is linear with applied field up to 9 T and shows almost no temperature dependence, indicating that hole carriers overwhelmingly dominate the electrical transport. This is indeed consistent with the theoretical prediction that only one Fermi pocket exists at the FL \cite{caagasdirac}. Figure 3(b) shows the MR of CaAgAs for some representative temperatures. While quadratic at small magnetic fields, it develops a linear behavior at high fields. Linear MR has also been observed in topological semimetals such as Na$_3$Bi \cite{na3bilinear} and Cd$_2$As$_3$ \cite{transcd3as2, linearcd3as2} with linear energy-momentum dispersions, as well as materials with parabolic dispersion, such as Ag$_{2-\delta}$Se \cite{Ag2Se} and GaAs quantum well \cite{GaAs}. Despite being a subject of study for decades, its origin is still under debate \cite{GaAs, linearMR}. Using $n=B/e\rho_{yx}$ and $\mu=R_H \sigma_{xx}$, the temperature dependent carrier density $n$ and mobility $\mu$ were estimated and shown in Fig. 3(c). $n$ is temperature independent and is of the order of $1.7 \times 10^{26}$ $\rm m^{-3}$ while $\mu$ shows a strong temperature dependence ranging from 0.3 $\rm{m^2V^{-1}s^{-1}}$ at 2 K, to 0.025 $\rm{m^2V^{-1}s^{-1}}$ at 300 K.

\begin{figure}
 \includegraphics[width=3.4in]{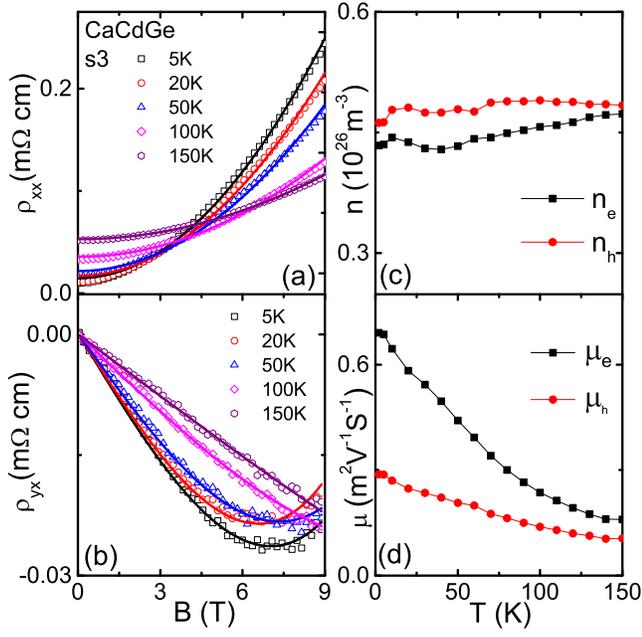}
  \caption{ CaCdGe single crystal s3 with $ I // c$ and B $\perp ac$: (a) Transverse magneto-resistivity $\rho_{xx}$. (b) Hall resistivity $\rho_{yx}$. The symbols correspond to experimental data, while the lines are the curves obtained from the two band model fitting. (c) Temperature dependent carrier densities. (d) Temperature dependent mobilities.}
 \label{fig:Fig1}
\end{figure}
\begin{figure}
 \includegraphics[width=3.4in]{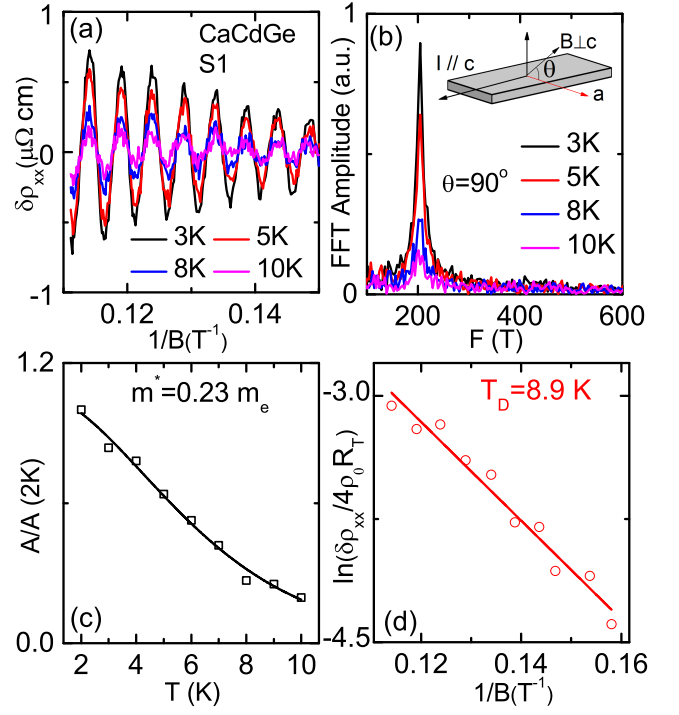}
\caption{(a) The oscillating part of $\rho_{xx}$, $\delta\rho_{xx}$, vs. 1/B. (b) FFT spectrum of $\delta\rho_{xx}$ for a few representative temperatures. Inset : Measurement configuration. The magnetic field was normal to the $ac$ plane. (c) Temperature dependence of the normalized amplitude of $\delta\rho_{xx}$ denoted as A/A(2K). (d) 1/B dependence of the quantity $ln(\delta\rho_{xx}/4\rho_0R_T)$, where $R_T= \frac{\alpha Tm^*/B}{sinh(\alpha Tm^*/B)}$, used to extract the Dingle temperature.}
\label{fig:Fig1}
\end{figure}

\begin{figure*}
\includegraphics[width=6.2in]{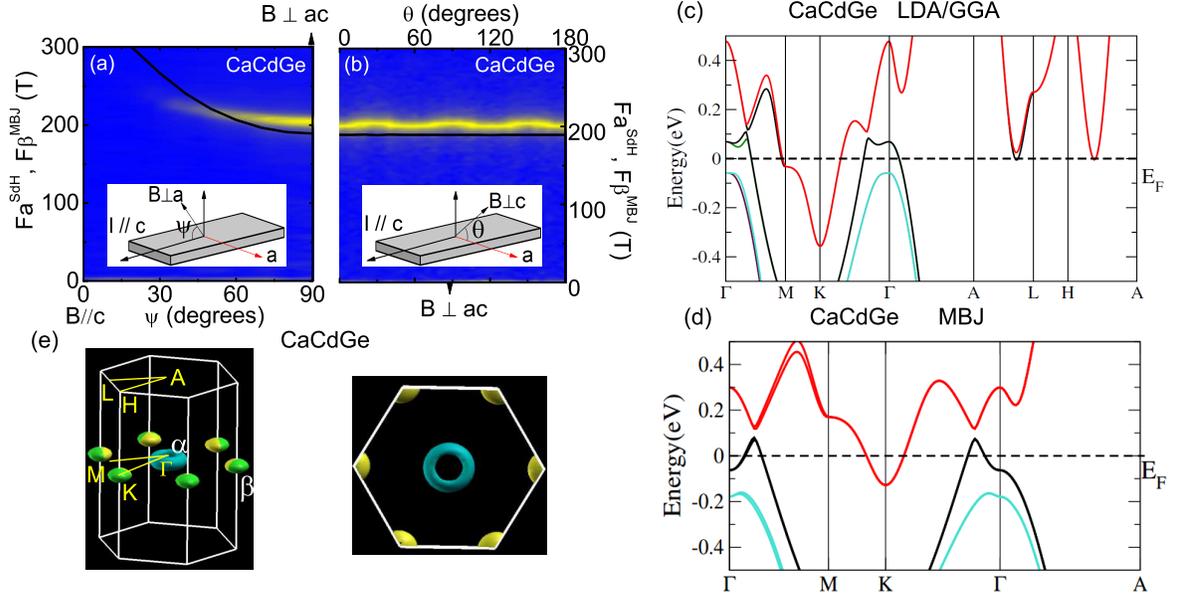}
 \caption{(a, b) Angular dependence of the experimental $F_{a}^{SdH}$ (yellow lines; see text) and the calculated $F_{\beta}^{SdH}$ (black lines; see text) with the measurement geometries in the insets. (c, d) The electronic band structure of CaCdGe with SOC: (c) Using the LDA/GGA potential. (d) Using the MBJ potential. (e) The Fermi pockets associated with (d). }
 \label{fig:Fig1}
\end{figure*}

As presented in Fig. 2(c), CaAgAs and CaCdGe show quite different behavior under magnetic field. In order to understand the origin of the large non-saturating MR observed in CaCdGe, we performed temperature dependent magnetotransport measurements. As a representative, the data measured on S3 with I // c and B $\perp ac$ is shown in Fig. 4. The transverse magnetoresistivity has a quadratic dependence on magnetic field (Fig. 4(a)). The pattern of the Hall resistivity $\rho_{yx}$ evolves with temperature and deviates from linear behavior at temperatures below 100 K (Fig. 4(b)). This suggests that transport in this compound is governed by both electron and hole carriers. To get a quantitative estimation of the carrier densities and mobilities in CaCdGe, we used the semiclassical two band model to analyze the data \cite{book, nbas2}. The field dependencies of $\rho_{xx}$ and $\rho_{yx}$ are given by
 \begin{equation}
 \rho_{xx} = E_x/J_x=\frac{n_e \mu_e + n_h\mu_h + (n_e\mu_h +n_h\mu_e)\mu_e\mu_hB^2}{e(n_e\mu_e + n_h\mu_h)^2 + e(n_h-n_e)^2\mu_e^2\mu_h^2B^2}
 \end{equation}
 and
 \begin{equation}
\rho_{yx} = E_y/J_x=\frac{B(n_h\mu_h^2 - n_e\mu_e^2) + (n_h - n_e)\mu_e^2\mu_h^2B^3}{e(n_e\mu_e + n_h\mu_h)^2 + e(n_h-n_e)^2\mu_e^2\mu_h^2B^2}
 \end{equation}
 where $n_e$, $n_h$, $\mu_e$ and $\mu_h$ are the fitting parameters, representing the carrier densities and mobilities of electrons and holes respectively. Simultaneously fitting our data with Eq. (1) and (2) allows us to determine the temperature dependence of $n_e$, $n_h$, $\mu_e$ and $\mu_h$. The fitting curves are presented as lines in Fig. 4 (a) and (b) and agree well with the experimental data. Figures 4 (c) and (d) show the temperature dependent $n_e$, $n_h$, $\mu_e$ and $\mu_h$ inferred from the fits. Both values of $n_e$ and $n_p$ are very similar to each other, being around $n \sim 5 \times \rm{10 ^{25}/m^3}$ and showing almost no temperature dependence. This suggests that the large non-saturating MR comes from the electron-hole compensation effect. Both $\mu_e$ and $\mu_h$ (Fig. 4(d)) increase with decreasing temperatures, being consistent with the weaker scattering at lower temperatures. At 2 K, $\mu_e$ is 0.7 $\rm{m^2/V/S}$ while $\mu_h$ is 0.3 $\rm{m^2/V/S}$.

In CaCdGe, Shubnikov-de Haas (SdH) oscillations were observed above 6 T in $\rho_{xx}$, as shown in Fig. 2(c). These oscillations are a result of the singularity in the density of states at the FL that occurs every time a Landau level crosses the FL. We analyzed the SdH data by first subtracting a polynomial background and then plotting $\delta\rho_{xx} = \rho_{xx} - \rho_{bkg}$ as a function of 1/B.  Figure 5(a) shows $\delta \rho_{xx}$ for S1 as a function of 1/B at a few representative temperatures with I // $c$ and B$\perp ac$. The oscillations are periodic in 1/B. Their frequency $F$ is related to the extremal cross sectional area $S$ of the Fermi surface perpendicular to the magnetic field through the Onsager relation $F = \hbar S/2\pi e$ \cite{book}. The Fast Fourier Transform (FFT) spectrum of the oscillations reveals only one frequency around 204 T, which is labelled as $F_a^{SdH}$ and shown in Fig. 5 (b). The amplitude of the oscillations, taking finite temperature and impurity scattering effects into account, is described by the Lifshitz-Kosevich formula
 \begin{equation}
\frac{\delta\rho_{xx}}{4\rho_0}=exp(-\alpha T_Dm^*/Bm_e)\frac{\alpha Tm^*/Bm_e}{sinh(\alpha Tm^*/Bm_e)}
 \end{equation}
where T$_D=\hbar/(2\pi k_B\tau)$ is the Dingle temperature, related to the quantum lifetime $\tau_q = \hbar / 2\pi k_BT_D$,  $\alpha = 2\pi^2k_Bm_e / \hbar = 14.69$ T/K, $m^*$ is the effective mass and $m_e$ the electron mass \cite{book}.

 Using this formalism, we extracted an effective mass $m^*$ of $0.23 m_e$ (Fig. 5(c)). Based on the Onsager relation, we estimated the Fermi wavevector $k_F=\sqrt{2eF/\hbar}$ to be $0.079  \AA^{-1}$ and the Fermi velocity $v_F=\hbar k_F/m^*$ as $4.0 \times10^5 $ m/s. Assuming the oscillations arise from an isotropic Fermi pocket, the carrier density $n=k_F^3/(3\pi^2)$ was estimated as $1.7 \times 10^{25}$ $\rm m^{-3}$. This value is 2- 3 times smaller than the ones inferred from Hall and MR data (Fig. 4(c)), suggesting that the associated Fermi surface is not of spherical shape. Our fit also resulted in a $T_D \rm{(2K)}=8.9 $ K (Fig. 5(d)) and thus a quantum lifetime $\tau_q \rm{(2K)}= 1.4  \times 10^{-13}$ s. Since $\tau_q$ arises from all scattering channels which broaden the Landau levels, while the transport lifetime $\tau_{tr}=m^*\mu/e$ mainly arises from backscattering, to investigate if there is a strong suppression of backscattering as is the case for Cd$_3$As$_2$ \cite{transcd3as2}, we compared $\tau_{q}$ and $\tau_{tr}$. Our estimation of $\tau_{tr}$ is quite rough due to the existence of two types of charge carriers and multiple Fermi pockets. Using the $m^*$ and the average of $\mu_e$ and $\mu_h$, the estimated $\tau_{tr}$ is around 6.5 $\times 10^{-13}$ s and the ratio of $\tau_{tr}/\tau_q$ is around 4.6. This ratio is significantly smaller than $10^4$ in Cd$_3$As$_2$ \cite{transcd3as2}.

To map out the electronic structure of CaCdGe experimentally, we studied the angle-dependent SdH oscillations. Figures 6 (a) and (b) present the angular dependence of $F_a^{SdH}$, with the measurement geometries shown in the insets. $F_a^{SdH}$ shows a 6-fold rotational symmetry with the ratio between the maxima and minima being 1.028 (Fig. 6 (b)). This is consistent with its hexagonal crystal structure and indicates small in-plane anisotropy of the associated Fermi pocket.

Figures 6(c) and (d) present the calculated band structures of CaCdGe using the LDA/GGA and MBJ potentials respectively. Regardless of the choice of potential, an extra trivial band (red), besides the nontrivial band carrying the nodal-line feature as proposed in CaAgAs (black), also crosses the FL in CaCdGe \cite{caagasdirac} . Therefore, both calculations support the electron-hole compensation effect suggested in Fig. 4(c) and (d). However, although Fig. 6(c) and (d) are quite similar in general, they differ in the details, which results in significant differences in the size and topology of the Fermi pockets. We were unable to relate $F_a^{SdH}$ with any Fermi pockets arising from the LDA/GGA band structure shown in Fig. 6(c) since all pockets are very large. On the other hand, the size of $F_a^{SdH}$ matches well with the ovoid-like $\beta$ Fermi pocket in Fig. 6(e), which originates from the trivial red band when the MBJ potential is used. The comparison between the $F_a^{SdH}$ (yellow lines) and the oscillation frequency associated with the $\beta$ pocket, $F_{\beta}^{SdH}$ (black lines) is shown in Fig. 6(a) and (b). We notice that although the magnitudes of both frequencies match quite well, $F_{\beta}^{SdH}$ shows stronger anisotropy in the $ac$ plane but weaker anisotropy in the $ab$ plane. Nevertheless, our observation suggests that the MBJ potential gives a better description of the band structure of CaCdGe. This may also be true for CaAgAs. Up to 9 T, we did not observe frequencies associated with the donut-like $\alpha$ Fermi pocket which originates from the band featuring the nodal-line. Higher magnetic field may be needed to fully unravel the Fermiology of CaCdGe.

\section{Conclusion}
In conclusion, we have grown and characterized the single crystalline Dirac nodal-line semimetal candidates CaAgAs and CaCdGe. Magnetoresistance measurements, Hall measurements and first-principles calculations indicate that CaAgAs is a single band material with one donut-like hole Fermi pocket, consistent with the proposal of being the first "hydrogen atom" nodal-line semimetal. First-principles calculations with the MBJ potential show that CaCdGe has one donut-like hole Fermi pocket originating from the band with the nodal-line feature and one trivial ovoid-like electron Fermi pocket. As a result, their magnetotransport properties behave quite differently. At 2 K and 9 T, linear transverse magnetoresistance (MR) up to 18\% is observed in CaAgAs while extremely large non-saturating quadratic MR up to 3200\% appears in CaCdGe, suggesting that the electron-hole compensation effect is responsible for the extremely large MR observed in CaCdGe. Angle-dependent SdH oscillations in CaCdGe resolve a Fermi pocket with oscillation frequency of 204 T and effective mass of 0.23 $m_e$, which agrees well with the ovoid-like hole Fermi pocket revealed by the first-principles calculations with the MBJ potential.

\section{Acknowledgments}
Work at UCLA was supported by the U.S. Department of Energy (DOE), Office of Science, Office of Basic Energy Sciences under Award Number DE-SC0011978. Work at Rutgers was supported by the NSF DMREF program under the award NSF DMREF project DMR-1435918. T. R. C. is supported by the Ministry of Science and Technology and National Cheng Kung University, Taiwan. T. R. C. also thanks National Center for Theoretical Sciences (NCTS), Taiwan for technical support. We thank Chang Liu for useful discussions.

\end{document}